\documentclass[11pt,a4paper]{article}
\pdfoutput=1
\usepackage{graphicx}
\usepackage{amsfonts}
\usepackage[margin=2cm]{geometry}
\usepackage{amsmath,amssymb,amsthm,array,booktabs,cite,dsfont,mathtools,multirow,placeins,kbordermatrix,rotating,stmaryrd,tensor,verbatim,subfig}
\usepackage[bookmarksnumbered,linktocpage,pdfstartview=FitH]{hyperref}
\usepackage{dcolumn}
\usepackage[all]{hypcap}

\usepackage{bm}
\usepackage{cancel}
\usepackage{amscd,slashed}

\newcolumntype{M}[1]{>{$}{#1}<{$}}

\DeclarePairedDelimiter{\ket}{\lvert}{\rangle}

\newcommand{\sst}[1]{{\scriptscriptstyle #1}}

\newcommand{\rep}[1]{\ensuremath{\mathbf{#1}}}

\def\0{{\sst{(0)}}}
\def\1{{\sst{(1)}}}
\def\2{{\sst{(2)}}}
\def\3{{\sst{(3)}}}
\def\4{{\sst{(4)}}}
\def\5{{\sst{(5)}}}
\def\6{{\sst{(6)}}}
\def\7{{\sst{(7)}}}

\newcommand{\be}{\begin{equation}}
\newcommand{\ee}{\end{equation}}
\def\ba{\begin{array}}
\def\ea{\end{array}}

\newtheorem*{theorem*}{Theorem}

\numberwithin{equation}{section}

\newcommand\half{\tfrac{1}{2}}

\newcommand{\bea}{\begin{eqnarray}}
\newcommand{\eea}{\end{eqnarray}}

\DeclareMathOperator{\SO}{SO}

\DeclareMathOperator{\USp}{USp}

\DeclareMathOperator{\SU}{SU}
\DeclareMathOperator{\Sp}{Sp}

\DeclareMathOperator{\Un}{U}

\DeclareMathOperator{\UOSp}{UOSp}
\DeclareMathOperator{\OSp}{OSp}
\DeclareMathOperator{\Spin}{Spin}
\DeclareMathOperator{\GL}{GL}

\newcommand{\R}{\mathds{R}}
\newcommand{\C}{\mathds{C}}
\newcommand{\Q}{\mathds{H}}

\DeclareMathOperator{\Ber}{Ber}

\begin{document}

\begin{titlepage}
\begin{center}
\hfill Imperial/TP/2014/mjd/04\\

\vskip 2cm

{\Huge \bf The structure of superqubit states}

\vskip 1.5cm

{\bf  L.~Borsten$^1$, K.~Br\'adler$^2$ and M.~J.~Duff$^1$}

\vskip 20pt

$^1${\it Theoretical Physics, Blackett Laboratory, Imperial College London,\\
 London SW7 2AZ, United Kingdom}\\\vskip 2pt
 $^2${\it Department of Astronomy and Physics,
Saint Mary's University,\\ Halifax, Nova Scotia, B3H 3C3, Canada}\\\vskip 5pt

\texttt{leron.borsten@imperial.ac.uk}\\
\texttt{kbradler@ap.smu.ca}\\
\texttt{m.duff@imperial.ac.uk}\\

\end{center}

\vskip 2.2cm

\begin{center} {\bf ABSTRACT}\\[3ex]\end{center}
Superqubits provide a supersymmetric generalisation of the conventional qubit in quantum information theory.  After a review of their current status, we address the problem of generating entangled states. We introduce the global unitary supergroup $\UOSp((3^n+1)/2 | (3^n-1)/2)$ for an $n$-superqubit system, which contains as a subgroup the local unitary supergroup $[\UOSp(2|1)]^n$. While for  $4>n>1$  the bosonic subgroup in $\UOSp((3^n+1)/2 | (3^n-1)/2)$ does not contain the standard global unitary group $\SU(2^n)$, it does have an $\USp(2^n)\subset\SU(2^n)$ subgroup which acts transitively on the  $n$-qubit subspace, as required for consistency with the conventional multi-qubit framework. For two superqubits the $\UOSp(5|4)$ action is used to generate entangled states from the ``bosonic'' separable state $\ket{00}$. 



\vfill


\end{titlepage}

\newpage \setcounter{page}{1} \numberwithin{equation}{section} \tableofcontents 

\newpage
\section{Introduction}

Superqubits \cite{Borsten:2009ae, Borsten:2012pp, Bradler:2012ii} belong to a $(2|1)$-dimensional complex super-Hilbert space \cite{Rogers:1979vp, DeWitt:1984, rudolph-2000-214}. In this sense they constitute the minimal supersymmetric generalisation of the conventional qubit. The use of super-Hilbert spaces implies a non-trivial departure from standard quantum theory; the term supersymmetry in this context should not be confused with its more familiar usage in the field of high energy physics, which operates exclusively in the realm of standard quantum theory. While the infinitesimal symmetries and  representations relevant to superqubits  appear in a variety of physically motivated contexts \cite{Wiegmann:1988, Sarkar:1991,  Essler:1992nk, Mavromatos:1999xj, Hasebe:2004hy,  Hasebe:2004yp, Arovas:2009dx, Hasebe:2011uq, Hasebe:2011ke, Hasebe:2013pra}, it is not clear what relation they have to superqubits themselves, especially given the use of super-Hilbert space.

Here we review  superqubits  and  develop further the  basic formalism,  paying particular attention to the issue of generating entangled states using global super-unitary transformations. In \autoref{sec:qubits} we briefly recall the essential features of conventional qubits. In \autoref{sec:super} we review in some detail Grassmann numbers, supermatrices, supergroups and Lie superalgebras. Having provided the necessary background, in \autoref{sec:superqubit} we review the superqubit formalism and introduce for the first time the super-unitary group acting globally on multi-superqubit states. It shown that this group  generates superentangled states.

Recall, a collection of $n$ distinct isolated qubits transforms under the group of local unitaries $[\SU(2)]^{ n}$. The local unitary transformations, by construction, cannot generate entanglement; a separable $n$-qubit state will remain separable under all $[\SU(2)]^{ n}$ operations. To generate an arbitrary state, entangled or otherwise, from any given initial state one conventionally  employs the group of global unitaries $\SU(2^n)$, which  acts transitively on the $n$-qubit state space. 

Similarly, a collection of $n$ distinct isolated superqubits transforms under the  local unitary orthosymplectic supergroup $[\UOSp(2|1)]^{ n}$, which contains as its bosonic subgroup the conventional local unitaries $[\SU(2)]^{ n}$. Again, being local, this set of transformations is insufficient to generate super-entanglement \cite{Borsten:2009ae}. With this issue in mind we introduce here the  $n$-superqubit  global unitary supergroup given by $\UOSp((3^n+1)/2 | (3^n-1)/2)$, the super-analog of  $\SU(2^n)$,  which is  uniquely determined by the single superqubit limit.

 At first sight this appears to present a  conundrum. For consistency  the  bosonic subgroup of $\UOSp((3^n+1)/2 | (3^n-1)/2)$ is required to act transitively on the subspace of regular qubits sitting inside the super-Hilbert space. However, for  $4>n>1$ the standard group of global unitaries $\SU(2^n)$ is \emph{not} contained in the bosonic subgroup of $\UOSp((3^n+1)/2 | (3^n-1)/2)$. Its absence, it would seem,   obstructs the expected consistent reduction to standard qubits. All is not lost however, since a proper subgroup $\USp(2^n)\subset\SU(2^n)$ is sufficient to generate any state from any initial state. This smaller group is indeed always contained in the bosonic subgroup of $\UOSp((3^n+1)/2 | (3^n-1)/2)$ and, as we shall explain, acts transitively on the subspace of standard $n$-qubit states. 
This is related  to the observation  made in the context of quantum control \cite{Albertini:2003} that the global unitary group carries a degree of redundancy and a $\USp(2^n)$ subgroup is always sufficient to generate all states, entangled or otherwise.  

\section{Qubits and global unitary groups}\label{sec:qubits}

Before turning the case of superqubits let us review the familiar case of regular qubits. 
The complex projective $n$-qubit state space 
\be
\mathds{P}(\C^2\otimes\ldots\otimes \C^2)\cong\C\mathds{P}^{2^n-1}\cong\SU(2^n)/\Un(2^n-1)\cong S^{2^{n+1}-1}/\Un(1)
\ee
is acted on transitively and effectively by  $\SU(2^n)$. As observed in the context of quantum control \cite{Albertini:2003} there is a proper subgroup $\USp(2^n)\subset\SU(2^n)$, which also acts transitively on the state space $\C\mathds{P}^{2^n-1}$, since
\be
\USp(2^n)/[\Un(1)\times\USp(2^n-2)]\cong\C\mathds{P}^{2^n-1}
\ee
via the  identification $\Q^m\cong\C^{2m}$. 

Neglecting the $\Un(1)$ quotient group and so distinguishing normalised states with distinct phase, yields $S^{2^{n+1}-1}$. The possible transitive actions on spheres were classified in \cite{Montgomery:1943,Borel:1949} and are summarised in \autoref{tab:spheres} (see also \cite{Besse:1987}).  \autoref{tab:spheres} presents a number of possible  $\Un(2^n)$ subgroups acting transitively on spheres, but it is only  $\USp(2^n)$  that is relevant to the case of finite-dimensional quantum systems described by finite-dimensional bilinear models \cite{Albertini:2003}. 

The conclusion of these observations is that any final $n$-qubit state, entangled or otherwise, may be obtained from any initial $n$-qubit state using only the $\USp(2^n)$ subgroup of the familiar unitary group $\Un(2^n)$. 

Note, while $\USp(2^n)\subset\SU(2^n)$ acts transitively on $\C\mathds{P}^{2^n-1}$, the group of local unitaries appropriate to the case of isolated qubits, 
\be
\SU_1(2)\times\ldots\times\SU_n(2)\subset \SU(2^n),
\ee is not a subgroup of $\USp(2^n)$. Although there is an $[\SU(2)]^{n}$ subgroup in $\USp(2^n)$ it cannot necessarily be identified with the local unitaries for $n>1$. This is most easily seen at $n=2$, for which we have the following branching 
\be
\begin{array}{cccccc}
\SU(4)&\supset& \USp(4)& \supset& \SU(2)\times\SU(2)\\
\mathbf{4}&\rightarrow&\mathbf{4}& \rightarrow& \mathbf{(2,1)\oplus(1,2)}\\
\end{array}
\ee
The $\SU(2)\times\SU(2)$ subgroup in  $\USp(4)$ is unique (up to conjugation) and therefore cannot be identified with the local unitaries $\SU_A(2)\times\SU_B(2) \subset \SU(4)$ since 
\be
\begin{array}{cccccc}
\SU(4)&\supset& \SU_A(2)\times\SU_B(2)\\
\mathbf{4}&\rightarrow& \mathbf{(2,2)}\\
\end{array}
\ee

  \begin{table}[ht]
 \begin{center}
\begin{tabular}{|c|c|c|c|}
\hline
Isometry  group $G$ & Sphere & Stabiliser group $K$ & $m$   \\
 \hline
 $\SO(n)$& $S^{n-1}$& $\SO(n-1)$& $0$\\
 \hline
 $\Un(n)$&  \multirow{2}{*}{$S^{2n-1}$}& $\Un(n-1)$ &1\\
 $\SU(n)$&& $\SU(n-1)$&1\\
 \hline
 $\USp(n)\times\USp(2)$&\multirow{3}{*}{$S^{4n-1}$}& $\USp(n-1)\times\USp(2)$ & $1$\\
 $\USp(n)\times\Un(1)$&& $\USp(n-1)\times\Un(1)$& $2$\\
 $\USp(n)$&& $\USp(n-1)$&$6$\\
 \hline
 $G_2$&$S^{6}$& $\SU(3)$&$0$\\
 $\Spin(7)$&$S^{7}$& $G_2$&$0$\\
  $\Spin(9)$&$S^{15}$& $\Spin(7)$&$1$\\
   \hline
\end{tabular}
\caption{Transitive actions on spheres. Here $K$ denotes the isotropy subgroup and $m$ indicates the dimension of the space of $G$-invariant Riemannian metrics up to homotheties. \label{tab:spheres}}
 \end{center}
\end{table}

\section{Supergroups}\label{sec:super}

\subsection{Grassmann numbers}

 Grassmann numbers are  elements of the Grassmann algebra $\Lambda_n$ over $\C$ (or $\R$ with analogous definitions), which is generated by $n$ mutually anticommuting elements $\{\theta^i\}_{i=1}^n$.

Any Grassmann number $z$ may be decomposed into ``body'' $z_{B}\in\mathds{C}$ and ``soul'' $z_{S}$ viz.
\begin{equation}
z=z_{B}+z_{S},\quad\text{where}\quad
z_{S}=\textstyle\sum_{k=1}^\infty\tfrac{1}{k!}c_{a_1\cdots a_k}\theta^{a_1}\cdots\theta^{a_k},
\end{equation}
and $c_{a_1\cdots a_k}\in\mathds{C}$ are totally antisymmetric. For finite dimension $n$ the sum terminates at $k=2^n$ and the soul is nilpotent $z_{{S}}^{n+1}=0$. One can take the formal limit $n\rightarrow \infty$, in which case elements of the algebra are often refereed to as supernumbers. 

One may also decompose $z$ into even and odd parts $u\in\Lambda_{n}^{0}$ and $v\in\Lambda_{n}^{1}$
\begin{equation}
\begin{split}
u&=z_\mathcal{B}+\textstyle\sum_{k=1}^\infty\tfrac{1}{(2k)!}c_{a_1\cdots a_{2k}}\theta^{a_1}\cdots\theta^{a_{2k}}\\
v&=\textstyle\sum_{k=0}^\infty\tfrac{1}{(2k+1)!}c_{a_1\cdots a_{2k+1}}\theta^{a_1}\cdots\theta^{a_{2k+1}},
\end{split}
\end{equation}
where $\Lambda_n=\Lambda_{n}^{0}\oplus\Lambda_{n}^{1}$.  For a Grassmann algebra over the complexes we also use $\C_c$ and $\C_a$ for the even commuting and odd anticommuting parts.

One defines the \emph{grade of a Grassmann number} as
\begin{equation}
\deg x:=\begin{cases}0&x\in\Lambda_{n}^{0}\\1&x\in\Lambda_{n}^{1},\end{cases}
\end{equation}
where the grades 0 and 1 are referred to as even and odd, respectively. Note, $xy=(-)^{xy}yx$ for $x,y \in \Lambda_{n}^{i}$. Here we have introduced the shorthand notation, $\deg \alpha\rightarrow\alpha$,  for any  $\deg \alpha$ appearing in the exponent of $(-)$. 

The superstar $^\#: \Lambda_n^i\rightarrow \Lambda_n^i$   is defined to satisfy,
\begin{equation}\label{eq:cond1}
(x \theta_i)^\#=x^* \theta_{i}^\#,\quad \theta_{i}^{\#\#}=-\theta_{i},\quad(\theta_i\theta_j)^\#=\theta_{i}^\#\theta_{j}^\#,
\end{equation}
where $x\in\mathds{C}$ and $^*$ is ordinary complex conjugation \cite{Scheunert:1977, Berezin:1981}. Hence,
\be\label{eq:cond2}
 \alpha^{\#\#}=(-)^{\alpha}\alpha
\ee
for pure even/odd Grassmann $\alpha$. The impure case follows by linearity.
\subsection{Supermatrices}

A $(p|q)\times(r|s)$ supermatrix is  a $(p+q)\times(r+s)$-dimensional block partitioned matrix
\begin{equation}
M= ~\kbordermatrix{&r&\vrule&s\\p&A&\vrule&B\\ \hline q&C&\vrule&D}
\end{equation}
with entries in a  Grassmann algebra, where supermatrix multiplication is defined as for ordinary matrices. Note, the special cases $r=1, s=0$ or $p=1, q=0$  correspond to  row and column supervectors. For notational convenience we will denote the  $(p|q)$-dimensional supervector space over a complex (real) Grassmann algebra by $\C^{p|q}$ ($\R^{p|q}$).

A supermatrix has a definite grade $\deg M=0,1$ if Grassmann entries in the $A$ and $D$ blocks are grade $\deg M$, and those in the $B$ and $C$ blocks are grade $\deg M+1\mod2$.  The supertranspose $M^{st}$ of a  supermatrix $M$ with $\deg M$ is defined componentwise as
\begin{equation}
M^{st}{}_{X_1X_2}:=(-)^{(X_1+M)(X_1+X_2)}M_{X_2X_1},
\end{equation}
where $X_1=1,\ldots p|p+1,\ldots p+q$ and $X_2=1,\ldots r|r+1,\ldots r+s$.  Note, we  have assigned supermatrix indices a grade in the obvious manner and addition in the exponent of $(-)$ is always mod 2.
 
In block matrix form
\be
\begin{bmatrix}A&B\\  C&D\end{bmatrix}^{st}=\begin{bmatrix}A^t&(-)^{M}C^t\\  (-)^{M+1}B^t&D^t\end{bmatrix},
\ee
so for column (row) vectors $V$  $(W)$ we have,
\be
V^{st}=\begin{bmatrix}x\\  y\end{bmatrix}^{st}=\begin{bmatrix}x^t&(-)^{V}y^t\end{bmatrix},\quad W^{st}=\begin{bmatrix}w&z\end{bmatrix}^{st}=\begin{bmatrix}w^t\\  (-)^{W+1}z^t\end{bmatrix}.
\ee
The supertranspose of an inhomogeneous grade supermatrix is defined by linearity. 

The supertranspose satisfies
\begin{equation}
\begin{split}
M^{st\,st}{}_{X_1X_2}&=(-)^{(X_1+X_2)}M_{X_1X_2},\\
M^{st\,st\,st}{}_{X_1X_2}&=(-)^{(X_2+M)(X_1+X_2)}M_{X_2X_1},\\
M^{st\,st\,st\,st}{}_{X_1X_2}&=M_{X_1X_2},
\end{split}
\end{equation}
and\begin{equation}
(MN)^{st}=(-)^{MN}N^{st}M^{st}.
\end{equation}

The  superadjoint $^\ddag$ of a supermatrix is defined as
\begin{equation}
M^\ddag:=M^{\# st},
\end{equation}
and satisfies
\begin{equation}
M^{\ddag\ddag}=(-)^{M}M,\quad (MN)^\ddag=(-)^{MN}N^\ddag M^\ddag.
\end{equation}

Note, the preservation of anti-super-Hermiticity, $M^\ddag=-M$, under scalar multiplication by Grassmann numbers necessitates the left/right multiplication rules \cite{buchbinder1998ideas},
\begin{equation}\label{eq:scalarmultiplication}
\begin{split}
(\alpha M)_{X_1X_2}&=(-)^{X_1\alpha}\alpha M_{X_1X_2},\\
(M\alpha)_{X_1X_2}&=(-)^{X_2\alpha}M_{X_1X_2}\alpha,
\end{split}
\end{equation}
or in block matrix form
\be\label{eq:inoutsign}
\alpha\begin{bmatrix}A&B\\  C&D\end{bmatrix}=\begin{bmatrix}\alpha A&\alpha B\\  (-)^{\alpha} \alpha C& (-)^{\alpha}\alpha D\end{bmatrix},\qquad \begin{bmatrix}A&B\\  C&D\end{bmatrix}\alpha=\begin{bmatrix} A\alpha&(-)^{\alpha} B\alpha \\    C \alpha & (-)^{\alpha} D\alpha\end{bmatrix}.
\ee

The  direct sum and super tensor product are unchanged from their ordinary versions up to the application of the sign rule in the commutatively isomorphism,
\be\label{eq:comm}
M\otimes N \mapsto (-)^{MN} N\otimes M,
\ee
 and  multiplication rule 
 \be
 (M_1\otimes N_1)(M_2\otimes N_2)=(-)^{N_1M_2}M_1M_2\otimes N_1N_2.
 \ee

\subsection{Orthosymplectic supergroups}
We will need to consider two isomorphic sets of Lie supergroups, 
\be
\OSp(2p+1|2q)\quad\text{and}\quad\OSp(2q | 2p+1),
\ee
 which are special cases of $\OSp(r|2q)$ and $\OSp(2q |r)$, respectively. See, for example,  \cite{Borsten:2009ae} and the references therein. 

As supermatrix groups they are  defined  as,
\be
\begin{split}
\OSp(2p+1|2q)&:=\left\{ X\in\GL(2p+1|2q)\, | \,  X^{st}(\eta\oplus\Omega) X= \eta\oplus\Omega \right\},\\[5pt]
\OSp(2q| 2p+1)&:=\left\{ X\in\GL(2q| 2p+1)\, | \, X^{st}(\Omega\oplus\eta) X= \Omega\oplus\eta \right\},\\
\end{split}
\ee
 where $\eta$ and $\Omega$ are the  symmetric and sympletic bilinear forms of $\SO(2p+1, \C)$ and $\Sp(2q, \C)$, respectively, and $\GL(r|s)$ denotes the supergroup of invertible  $(r|s)\times(r|s)$ even supermatrices. In the following we will only discuss  $\OSp(2p+1|2q)$ as the structure of $\OSp(2q|2p+1)$ trivially follows; one simply sends $\delta_{X_1X_4}\delta_{X_2X_3}$ to $\delta_{X_1X_3}\delta_{X_2X_4}$ in the definition of $U$ given in \eqref{eq:Gdef} below.

The $\mathfrak{osp}(2p+1|2q)$ superalgebra in the defining representation  can be  constructed using the supermatrices
\be\label{eq:Gdef}
(U_{X_1X_2})_{X_3X_4}:=\delta_{X_1X_4}\delta_{X_2X_3},\quad\text{and}\quad
G:=\begin{bmatrix}\eta&\vrule&0\\  \hline 0&\vrule&\Omega\end{bmatrix}.
\ee
 Here the indices $X_i$ range from 1 to $2p+1+2q$ and are partitioned as $X_i=(\mu, a)$ with $\mu$ ranging from 1 to $2p+1$, and $a$ taking on the remaining $2q$ values. 
 
The generators $T\in \mathfrak{osp}(2p+1|2q)$ are then given by,
\begin{equation}\label{eq:tdef}
T_{X_1X_2}=2G_{\llbracket X_1|X_3}U_{X_3|X_2 \rrbracket},
\end{equation}
where $T$ has array grade zero and we have introduced the graded symmetrization of superarrays, 
\begin{equation}
M_{X_1\cdots \llbracket X_i|\cdots|X_j\rrbracket\cdots X_k}:=
\half[M_{X_1\cdots X_i\cdots X_j\cdots X_k}+(-)^{(X_i+1)(X_j+1)}M_{X_1\cdots  X_j\cdots X_i\cdots X_k}].
\end{equation} Explicitly,
\be
\begin{split}
T_{\mu\nu}&=G_{\mu \lambda}U_{\lambda \nu}-G_{\nu \lambda}U_{\lambda \mu},\\
T_{ab}&=G_{a c}U_{c b}+G_{b c}U_{c a},\\
T_{\mu b}&=G_{\mu \lambda}U_{\lambda b}+G_{b c}U_{c \mu}.\\
\end{split}
\ee
Clearly $T$ has symmetry properties $T_{X_1X_2}=T_{\llbracket X_1X_2\rrbracket}$. The $2q(2q+1)/2$ elements $T_{a_1a_2}$ generate $\mathfrak{sp}(2q)$, the $(2p+1)(2p)/2$ elements $T_{\mu_1\mu_2}$ generate $\mathfrak{so}(2p+1)$, and both are even (bosonic), while the $(2p+1)2q$ generators $T_{\mu a}$ are odd (fermionic). These supermatrices yield the $\mathfrak{osp}(2p+1|2q)$ superbrackets
\begin{equation}
\llbracket T_{X_1X_2},T_{X_3X_4}\rrbracket:=4G_{\llbracket X_1\llbracket X_3}T_{X_2\rrbracket X_4\rrbracket},
\end{equation}
where the supersymmetrization on the right-hand side is over pairs $X_1X_2, X_3X_4$ as on the left-hand side and we have defined the  superbracket 
\begin{equation}
\llbracket M_{X_1X_2},N_{X_3X_4}\rrbracket:=
M_{X_1X_2}N_{X_3X_4}-(-)^{(X_1+X_2)(X_3+X_4)}N_{X_3X_4}M_{X_1X_2}.
\end{equation}
The action of the generators on $\C^{(2p+1|2q)}$, which constitutes a left $\mathfrak{osp}(2p+1|2q)$-supermodule, is given by
\begin{equation}
(T_{X_1X_2})_{X_3X_4}a_{X_4}\equiv(T_{X_1X_2}a)_{X_3}=2G_{\llbracket X_1|X_3}a_{X_2\rrbracket},\quad a\in \C^{2p+1|2q}.
\end{equation}
This action may be generalized to an $N$-fold super tensor product of $(2p+1|2q)$ supervectors by labeling the indices with integers $k=1,2,\dotsc,N$
\begin{equation}
(T_{X_kY_k}a)_{Z_1\cdots Z_k\cdots Z_N}=
(-)^{(X_k+Y_k)\sum_{i=1}^{k-1}|Z_i|}2G_{\llbracket X_k|Z_k}a_{Z_1\cdots |Y_k\rrbracket\cdots Z_N}.
\end{equation}

The Grassmann envelop or left supermodule $\mathfrak{osp}(2p+1|2q; \mathfrak{G})$ \cite{berezin1987introduction, Berezin:1981}  is given by the set of even supermatrices
\be
X=\xi_{X_1X_2}T_{X_1X_2}
\ee
where the $\xi_{X_1X_2}$ are even or odd complex Grassmann numbers if $\deg(X_1)+\deg(X_2)=0$ or $1$, respectively. The identity connected component of the supergroup $\OSp(2p+1|2q)$ is given by the exponential map of  $\mathfrak{osp}(2p+1|2q; \mathfrak{G})$.

\subsection{Unitary orthosymplectic supergroups}

As a supermatrix group the ``real form'' $\UOSp(2p+1|2q)$ is  defined  as,
\be
\UOSp(2p+1|2q):=\left\{ X\in\OSp(2p+1|2q)\, | \,  X^{\ddagger}=X^{-1},\quad \Ber(X)=1 \right\},
\ee
where 
\begin{equation}\label{eq:Ber}
\Ber M:=\det(A-BD^{-1}C)/\det(D)
      =\det(A)/\det( D - C A^{-1} B)
\end{equation}
is the Berezinian.

The corresponding Lie  algebra is given by,
\be
\mathfrak{uosp}(2p+1|2q; \mathfrak{G}):=\{X\in\mathfrak{osp}(2p+1|2q; \mathfrak{G}) | X^\ddag=-X\}.
\ee
Under supertransposition the $\mathfrak{osp}(2p+1|2q)$ generators obey
\be\label{eq:strelations}
T_{X_1X_2}{}^{st} = (-)^{X_1+X_2}G_{X_1X'_{1}}G_{X_2X'_{2}}T_{X'_1X'_2}
\ee
or in blocks
\be
T_{\mu\nu}{}^{st} = -\eta_{\mu\mu'}\eta_{\nu\nu'}T_{\mu'\nu'},\qquad
T_{a b}{}^{st} = -\Omega_{aa'}\Omega_{bb'}T_{a' b'},\qquad
T_{\mu b}{}^{st} = \eta_{\mu\mu'}\Omega_{bb'}T_{\mu' b'}.
\ee
Note, here we are taking the supertranspose of the supermatrices $T_{X_1X_2}$ defined by \eqref{eq:tdef}; the indices here label an  $X_1\times X_2$ array of supermatrices, not elements of a supermatrix. Using \eqref{eq:strelations}   even grade elements $X\in\mathfrak{uosp}(2p+1|2q; \mathfrak{G})$ can be written,
\be
X=(\alpha_{\mu\nu}+\eta_{\mu\mu'}\eta_{\nu\nu'}\alpha^{\#}_{\mu'\nu'})T_{\mu\nu}+(\beta_{a b}+\Omega_{a a'}\Omega_{bb'}\beta^{\#}_{a'b'})T_{ab},
\ee
where $\alpha, \beta$ are commuting Grassmann numbers. The  odd grade elements $X\in\mathfrak{uosp}(2p+1|2q; \mathfrak{G})$ are give by,
\be
X=(\tau_{\mu b}+\eta_{\mu \mu'}\Omega_{bb'}\tau^{\#}_{\mu' b'})T_{\mu b},
\ee
where $\tau$ are anticommuting Grassmann numbers.
Hence, a generic element $X\in\mathfrak{uosp}(2p+1|2q; \mathfrak{G})$ can be written,
 \be\label{eq:X}
 X=(\alpha_{\mu\nu}+\eta_{\mu\mu'}\eta_{\nu\nu'}\alpha^{\#}_{\mu'\nu'})T_{\mu\nu}+(\beta_{a b}+\Omega_{a a'}\Omega_{bb'}\beta^{\#}_{a'b'})T_{ab}+(\tau_{\mu b}+\eta_{\mu\mu'}\Omega_{bb'}\tau^{\#}_{\mu' b'})T_{\mu b}.
 \ee

\section{Superqubits  and global unitary supergroups}\label{sec:superqubit}

 The $n$-superqubit states are given by elements
\be\label{eq:nsuperqubits}
\ket{\psi}=a_{X_1\ldots X_2}\ket{X_1\ldots X_2},\qquad X_i= 0,1, \bullet
\ee
of  the $n$-fold tensor product super Hilbert space
\be\label{eq:hilb1}
\C_{1}^{2|1}\otimes\ldots\otimes\C_{n}^{2|1},
\ee
which constitutes the fundamental representation of the super local operations group $[\UOSp(2|1)]^{\otimes n}$.

The  supergroup  of global unitary operations is given by
\be\label{eq:nluosp}
\UOSp(\tfrac{3^n+1}{2} | \tfrac{3^n-1}{2})
\ee
acting on the $3^n$-dimensional graded vector space 
\be
\C^{\tfrac{3^n+1}{2} | \tfrac{3^n-1}{2}}\cong \C_{1}^{2|1}\otimes\ldots\otimes\C_{n}^{2|1}.
\ee
Note, here we have reordered the alternating even/odd grades inherited from the tensor product into to the standard $\tfrac{3^n+1}{2} | \tfrac{3^n-1}{2}$ convention.

The global supergroup \eqref{eq:nluosp} has body,
\be
\USp(\tfrac{3^n+1}{2}) \times \SO(\tfrac{3^{n}-1}{2}),\qquad\SO(\tfrac{3^{n}+1}{2})\times \USp(\tfrac{3^n-1}{2}),
\ee
for $n=2m+1$ and $n=2m$, respectively. Under the bosonic subgroup \eqref{eq:nsuperqubits} transforms in the defining representation
\be\label{eq:defrep}
V_n\otimes \mathbf{1}\oplus \mathbf{1} \otimes W_n, \qquad W_n\otimes \mathbf{1}\oplus \mathbf{1} \otimes V_n,
\ee
for $n=2m+1$ and $n=2m$, respectively, where $V_n$  and $W_n$ denote the defining vector and  symplectic vector representations of  $\SO$ and $\USp$. Under the  $[\SU(2)]^{n}$ body of  $[\UOSp(2|1)]^{n}\subset\UOSp(\tfrac{3^n+1}{2} | \tfrac{3^n-1}{2})$ the $n$-superqubit representation \eqref{eq:defrep} branches to
\be
\bigoplus_{p=1}^{n}\left[ \bigoplus_{\text{distinct permutations}}^{{{}^nC_{p}}}(\, \overbrace{\mathbf{2},\mathbf{2},\ldots , \mathbf{2}}^{p}, \underbrace{\mathbf{1}, \mathbf{1},\ldots , \mathbf{1}}_{n-p}\,) \right].
\ee
The even (odd) graded subspaces are  given by $n-p$ even (odd).

\subsection{Two superqubits: $\UOSp(5 | 4)\supset \UOSp_A(2|1)\times\UOSp_B(2|1)$}
Two superqubit states 
\be\label{eq:2superqubits}
\ket{\psi}=a_{XY}\ket{X Y}=a_{AB}\ket{AB}+a_{A\bullet}\ket{A\bullet}+a_{\bullet B}\ket{\bullet B}+a_{\bullet\bullet}\ket{\bullet\bullet}
\ee
transforms as the 
\be
\mathbf{(5,1)\oplus (1,4)}
\ee
of $\SO(5)\times\USp(4)$, where the even subspace spanned by $\{\ket{AB}, \ket{\bullet\bullet}\}$ constitutes the $\mathbf{(5,1)}$ and the odd subspace spanned by $\{\ket{A\bullet}, \ket{\bullet B}\}$ constitutes the $\mathbf{(1,4)}$.

There is a unique $\SU(2)\times\SU(2)$ in $\SO(5)\cong \UOSp(4)$ under which 
\be\label{eq:5-22}
\mathbf{5}\rightarrow \mathbf{(1,1)\oplus (2,2)},\qquad \mathbf{4}\rightarrow \mathbf{(2,1)\oplus (1,2)}.
\ee
Hence, there is a diagonal $\SU_A\times\SU_B$ in $[\SU(2)]^{4}\subset\SO(5)\times \UOSp(4)$ under which 
\be
\mathbf{(5,1)\oplus (1,4)} \rightarrow \mathbf{(2,2)\oplus (2,1)}\oplus \mathbf{(1,2)\oplus (1,1)},
\ee
where the summands are spanned by $\{\ket{AB}\},\{\ket{A\bullet}\}, \{\ket{\bullet B}\}$ and  $\{\ket{\bullet\bullet}\}$, respectively.

Letting 
\be
\eta=\begin{bmatrix}\mathds{1}&0&0\\0&\mathds{1}&0\\0&0&1\end{bmatrix}, \quad \Omega=\begin{bmatrix}0&\mathds{1}\\-\mathds{1}&0\end{bmatrix},
\ee
a general element   $X\in\mathfrak{uosp}(5|4; \mathfrak{G})$ may be written,
\be
\tiny
X=\left[
\begin{array}{ccccc|cccc}
 0 & -\alpha _{12}^{\#}-\alpha _{12} & -\alpha _{13}^{\#}-\alpha _{13} &- \alpha
   _{14}^{\#}-\alpha _{14} &- \alpha _{15}^{\#}-\alpha _{15} & \tau _{11}^{\#}-\tau _{13} &
   \tau _{12}^{\#}-\tau _{14} & \tau _{13}^{\#}+\tau _{11} & \tau _{14}^{\#}+\tau _{12} \\[5pt] 
 \alpha _{12}+\alpha _{12}^{\#} & 0 & -\alpha _{23}^{\#}-\alpha _{23} & -\alpha
   _{24}^{\#}-\alpha _{24} &- \alpha _{25}^{\#}-\alpha _{25} & \tau _{21}^{\#}-\tau _{23} &
   \tau _{22}^{\#}-\tau _{24} & \tau _{23}^{\#}+\tau _{21} & \tau _{24}^{\#}+\tau _{22} \\[5pt] 
 \alpha _{13}+\alpha _{13}^{\#} & \alpha _{23}+\alpha _{23}^{\#} & 0 & -\alpha
   _{34}^{\#}-\alpha _{34} & -\alpha _{35}^{\#}-\alpha _{35} & \tau _{31}^{\#}-\tau _{33} &
   \tau _{32}^{\#}-\tau _{34} & \tau _{33}^{\#}+\tau _{31} & \tau _{34}^{\#}+\tau _{32} \\[5pt] 
 \alpha _{14}+\alpha _{14}^{\#} & \alpha _{24}+\alpha _{24}^{\#} & \alpha _{34}+\alpha
   _{34}^{\#} & 0 & -\alpha _{45}^{\#}-\alpha _{45} & \tau _{41}^{\#}-\tau _{43} & \tau
   _{42}^{\#}-\tau _{44} & \tau _{43}^{\#}+\tau _{41} & \tau _{44}^{\#}+\tau _{42} \\[5pt] 
 \alpha _{15}+\alpha _{15}^{\#} & \alpha _{25}+\alpha _{25}^{\#} & \alpha _{35}+\alpha
   _{35}^{\#} & \alpha _{45}+\alpha _{45}^{\#} & 0 & \tau _{51}^{\#}-\tau _{53} & \tau
   _{52}^{\#}-\tau _{54} & \tau _{53}^{\#}+\tau _{51} & \tau _{54}^{\#}+\tau _{52} \\[5pt] 
   \hline
   &&&&&&&\\
 -\tau _{13}^{\#}-\tau _{11} & -\tau _{23}^{\#}-\tau _{21} & -\tau _{33}^{\#}-\tau _{31} & -\tau
   _{43}^{\#}-\tau _{41} & -\tau _{53}^{\#}-\tau _{51} & \beta _{13}^{\#}-\beta _{13} & \beta
   _{23}^{\#}-\beta _{14} & \beta _{11}+\beta _{33}^{\#} & \beta _{12}+\beta _{34}^{\#}
   \\[5pt] 
 -\tau _{14}^{\#}-\tau _{12} & -\tau _{24}^{\#}-\tau _{22} & -\tau _{34}^{\#}-\tau _{32} & -\tau
   _{44}^{\#}-\tau _{42} & -\tau _{54}^{\#}-\tau _{52} & \beta _{14}^{\#}-\beta _{23} & \beta
   _{24}^{\#}-\beta _{24} & \beta _{12}+\beta _{34}^{\#} & \beta _{22}+\beta _{44}^{\#}
   \\[5pt] 
 -\tau _{13}+\tau _{11}^{\#} & -\tau _{23}+\tau _{21}^{\#} & -\tau _{33}+\tau _{31}^{\#} & -\tau
   _{43}+\tau _{41}^{\#} & -\tau _{53}+\tau _{51}^{\#} & -\beta _{11}^{\#}-\beta _{33} &- \beta
   _{12}^{\#}-\beta _{34} & \beta _{13}-\beta _{13}^{\#} & \beta _{23}-\beta _{14}^{\#}
   \\[5pt] 
 -\tau _{14}+\tau _{12}^{\#} & -\tau _{24}+\tau _{22}^{\#} & -\tau _{34}+\tau _{32}^{\#} & -\tau
   _{44}+\tau _{42}^{\#} & -\tau _{54}+\tau _{52}^{\#} &- \beta _{12}^{\#}-\beta _{34} & -\beta
   _{22}^{\#}-\beta _{44} & \beta _{14}-\beta _{23}^{\#} & \beta _{24}-\beta _{24}^{\#}
   \\[5pt] 
\end{array}
\right]
\ee
On setting soul  to zero we find
\be
X=\begin{bmatrix}A & \vrule &0&0\\ \hline 0 &\vrule& B & C \\0 &\vrule& -C & -B^{t} \end{bmatrix}
\ee
where $A$ is real and antisymmetric while $B^\dagger=-B$ and $C$ is complex symmetric, the Lie algebra of $\mathfrak{so}(5)\oplus\mathfrak{usp}(4)$.

The even grade subspace spanned by $\{\ket{AB}, \ket{\bullet\bullet}\}$ is acted on transitively by  $\SO(5)\subset\UOSp(5|4)$ as a real 5-dimensional representation. Now recall that every physical 2-qubit state in $\C\mathds{P}^3$ is equivalent under local unitaries $\SU_A(2)\times\SU_B(2)$ to  a real state:
\be
\ket{\psi}\rightarrow \cos \phi \ket{00}+\sin \phi \ket{11}.
\ee
Hence, the bosonic $\SO(5)\cong \USp(4)\subset \SU(4)$ subgroup of $\UOSp(5|4)$ acts transitively on the 2-qubit subspace, as required, even though it does not contain the full global unitary group $\SU(4)$. Note however, it follows from \eqref{eq:5-22} that this $\SO(5)\cong \USp(4)\subset \UOSp(5|4)$ cannot be identified with the $\USp(4)\subset \SU(4)$ of quantum control under which the 2-qubit state transforms as the  $\rep{4}$. 

\subsection{Generating entangled states}
To identify global $\UOSp(5|4)$ transformations not contained in the local $\UOSp_A(2|1)\times\UOSp_B(2|1)$ subgroup it is convenient to work in the 2-superqubit tensor product basis. Let 
\be
 \tilde{G} = g \otimes g,
 \ee
 where
 \be
g=\begin{bmatrix}\varepsilon &\vrule &0\\\hline 0 &\vrule &1\end{bmatrix}=\begin{bmatrix}0&1&\vrule &0\\-1&0&\vrule &0\\\hline 0&0 &\vrule &1\end{bmatrix}
\ee
is the $\mathfrak{osp}(2|1)$ invariant tensor for a single superqubit. Permuting the 2-superqubit basis states into the canonical $(5|4)$ basis using a similarity transformation 
\be
S: \begin{matrix}\ket{00}\\\ket{01}\\\ket{0\bullet}\\\ket{10}\\\ket{11}\\\ket{1\bullet}\\\ket{\bullet 0}\\\ket{\bullet1}\\\ket{\bullet\bullet}\end{matrix}\mapsto \begin{matrix}\ket{00}\\\ket{01}\\\ket{10}\\\ket{11}\\\ket{\bullet\bullet} \\ \ket{0\bullet}\\\ket{\bullet 0}\\\ket{1\bullet}\\\ket{\bullet1} \end{matrix}
\ee
gives
\be\label{eq:54basis}
G = S\tilde{G}S^T =\eta\oplus\Omega 
,\quad\text{where}\quad
\eta=\begin{bmatrix}0&\varepsilon &0\\-\varepsilon&0&0\\0&0&1\end{bmatrix}.
\ee
  Applying \eqref{eq:X} we obtain a super-anti-Hermitian  $X\in\mathfrak{uosp}(5|4; \mathfrak{G})$ preserving  \eqref{eq:54basis}:
\be\label{eq:X2121}
\tiny
\left[
\begin{array}{ccccc|cccc}
 \alpha _{14}^{\#}-\alpha _{14} & \alpha _{24}^{\#}+\alpha _{13} & \alpha _{34}^{\#}+\alpha
   _{12} & 0 & -\alpha _{45}^{\#}-\alpha _{15} & \tau _{41}^{\#}-\tau _{13} & \tau _{42}^{\#}-\tau
   _{14} & \tau _{43}^{\#}+\tau _{11} & \tau _{44}^{\#}+\tau _{12} \\[5pt] 
 -\alpha _{13}^{\#}-\alpha _{24} & \alpha _{23}-\alpha _{23}^{\#} & 0 & \alpha
   _{34}^{\#}+\alpha _{12} & \alpha _{35}^{\#}-\alpha _{25} & -\tau _{31}^{\#}-\tau _{23} & -\tau
   _{32}^{\#}-\tau _{24} & \tau _{21}-\tau _{33}^{\#} & \tau _{22}-\tau _{34}^{\#} \\[5pt] 
 -\alpha _{12}^{\#}-\alpha _{34} & 0 & \alpha _{23}^{\#}-\alpha _{23} & \alpha
   _{24}^{\#}+\alpha _{13} & \alpha _{25}^{\#}-\alpha _{35} & -\tau _{21}^{\#}-\tau _{33} & -\tau
   _{22}^{\#}-\tau _{34} & \tau _{31}-\tau _{23}^{\#} & \tau _{32}-\tau _{24}^{\#} \\[5pt] 
 0 & -\alpha _{12}^{\#}-\alpha _{34} & -\alpha _{13}^{\#}-\alpha _{24} & \alpha _{14}-\alpha
   _{14}^{\#} & -\alpha _{15}^{\#}-\alpha _{45} & \tau _{11}^{\#}-\tau _{43} & \tau
   _{12}^{\#}-\tau _{44} & \tau _{13}^{\#}+\tau _{41} & \tau _{14}^{\#}+\tau _{42} \\[5pt] 
 \alpha _{15}^{\#}+\alpha _{45} & \alpha _{25}^{\#}-\alpha _{35} & \alpha _{35}^{\#}-\alpha
   _{25} & \alpha _{45}^{\#}+\alpha _{15} & 0 & \tau _{51}^{\#}-\tau _{53} & \tau _{52}^{\#}-\tau
   _{54} & \tau _{53}^{\#}+\tau _{51} & \tau _{54}^{\#}+\tau _{52} \\[5pt] \hline
      &&&&&&&\\
 -\tau _{13}^{\#}-\tau _{41} & \tau _{31}-\tau _{23}^{\#} & \tau _{21}-\tau _{33}^{\#} & -\tau
   _{43}^{\#}-\tau _{11} & -\tau _{53}^{\#}-\tau _{51} & \beta _{13}^{\#}-\beta _{13} & \beta
   _{23}^{\#}-\beta _{14} & \beta _{33}^{\#}+\beta _{11} & \beta {34}^{\#}+\beta _{12} \\[5pt] 
 -\tau _{14}^{\#}-\tau _{42} & \tau _{32}-\tau _{24}^{\#} & \tau _{22}-\tau _{34}^{\#} & -\tau
   _{44}^{\#}-\tau _{12} & -\tau _{54}^{\#}-\tau _{52} & \beta _{14}^{\#}-\beta _{23} & \beta
   _{24}^{\#}-\beta _{24} & \beta _{34}^{\#}+\beta _{12} & \beta _{44}^{\#}+\beta _{22} \\[5pt] 
 \tau _{11}^{\#}-\tau _{43} & \tau _{21}^{\#}+\tau _{33} & \tau _{31}^{\#}+\tau _{23} & \tau
   _{41}^{\#}-\tau _{13} & \tau _{51}^{\#}-\tau _{53} & -\beta _{11}^{\#}-\beta _{33} & -\beta
   _{12}^{\#}-\beta _{34} & \beta _{13}-\beta _{13}^{\#} & \beta _{23}-\beta _{14}^{\#} \\[5pt] 
 \tau _{12}^{\#}-\tau _{44} & \tau _{22}^{\#}+\tau _{34} & \tau _{32}^{\#}+\tau _{24} & \tau
   _{42}^{\#}-\tau _{14} & \tau _{52}^{\#}-\tau _{54} & -\beta _{12}^{\#}-\beta _{34} & -\beta
   _{22}^{\#}-\beta _{44} & \beta _{14}-\beta _{23}^{\#} & \beta _{24}-\beta _{24}^{\#} \\[5pt] 
\end{array}
\right]
\ee
Using a straightforward reparametrisation of \eqref{eq:X2121} the  $\mathfrak{uosp}_A(2|1)\oplus\mathfrak{uosp}_B(2|1)$ subalgebra can be written as:
\be\label{eq:2121}
\tiny
\left[
\begin{array}{ccccc|cccc}
i\gamma+i\delta&
   \delta _{+}+i\delta _{-} & \gamma_++i\gamma_-& 0
   & 0 & -\sigma^{\#} & -\rho^{\#} & 0 & 0 \\[5pt] 
- \delta _{+}+i\delta _{-} &i\gamma-i\delta& 0 & \gamma_++i\gamma_- & 0 & -\sigma & 0 & 0 & -\rho^{\#} \\[5pt] 
 -\gamma_++i\gamma_- & 0 &-i\gamma+i\delta&  \delta _{+}+i\delta _{-} &
   0 & 0 & -\rho&-\sigma^{\#} & 0 \\[5pt] 
 0 & -\gamma_++i\gamma_-&  -\delta _{+}+i\delta _{-}&
- i\gamma-i\delta& 0 & 0
   & 0 & -\sigma & -\rho \\[5pt] 
 0 & 0 & 0 & 0 & 0 & -\rho & \sigma &
   \rho^{\#} & -\sigma^{\#} \\[5pt] 
   \hline
      &&&&&&&\\
 \sigma & -\sigma^{\#} & 0 & 0 & -\rho^{\#} &i\gamma & 0 & \gamma_++i\gamma_-& 0 \\[5pt] 
\rho & 0 & -\rho^{\#} & 0 & \sigma^{\#} & 0 & i\delta& 0 & \delta _{+}+i\delta _{-} \\[5pt] 
 0 & 0 & \sigma & -\sigma^{\#} &
   -\rho & -\gamma_++i\gamma_-& 0 & -i\gamma& 0 \\[5pt] 
 0 & \rho & 0 & -\rho^{\#} & \sigma & 0 & - \delta _{+}+i\delta _{-} & 0 &-i\delta \\[5pt] 
\end{array}
\right]
\ee
where $\gamma_{(\pm)}^{\#}=\gamma_{(\pm)}, \delta_{(\pm)}^{\#}=\delta_{(\pm)}$,  which  follows from the graded tensor product \eqref{eq:comm},
\be
S(x_A\otimes\mathds{1}+\mathds{1}\otimes x_B)S^T,
\ee
where
\be
\begin{array}{llllll}
x_A &= &\left[
\begin{array}{ccc}
 i\gamma & \gamma_++i\gamma_-   &
  -\rho^{\#}  \\[5pt] 
-\gamma_++i\gamma_-& -i\gamma &
  -\rho  \\[5pt] 
 \rho  &- \rho^{\#} & 0 \\[5pt] 
\end{array}\right]&\in&\mathfrak{uosp}_A(2|1),\\[35pt]
x_B&= &\left[
\begin{array}{ccc}
 i\delta & \delta_++i\delta_-   &
  -\sigma^{\#}  \\[5pt] 
-\delta_++i\delta_-& -i\delta &
  -\sigma  \\[5pt] 
 \sigma  &- \sigma^{\#} & 0 \\[5pt] 
\end{array}\right]&\in& \mathfrak{uosp}_B(2|1).
\end{array}
\ee
A particularly simple example of a super entangled state is given by  exponentiating  \eqref{eq:X2121}  with only $\tau_{12}\in\C_a$ non-zero:
\be\text{exp}(X_{\tau_{12}})=
\left[
\begin{array}{ccccccccc}
 1+\frac{1}{2} \tau _{12}\tau _{12}^{\#}  & 0 & 0 & 0 & 0 & 0 & 0 & 0 & \tau _{12} \\[5pt] 
 0 & 1 & 0 & 0 & 0 & 0 & 0 & 0 & 0 \\[5pt] 
 0 & 0 & 1 & 0 & 0 & 0 & 0 & 0 & 0 \\[5pt] 
 0 & 0 & 0 & 1+\frac{1}{2} \tau _{12} \tau _{12}^{\#}  & 0 & 0 & \tau _{12}^{\#} & 0 & 0 \\[5pt] 
 0 & 0 & 0 & 0 & 1 & 0 & 0 & 0 & 0 \\[5pt] 
 0 & 0 & 0 & 0 & 0 & 1 & 0 & 0 & 0 \\[5pt] 
 0 & 0 & 0 & -\tau _{12} & 0 & 0 & 1-\frac{1}{2} \tau _{12}\tau _{12}^{\#}  & 0 & 0 \\[5pt] 
 0 & 0 & 0 & 0 & 0 & 0 & 0 & 1 & 0 \\[5pt] 
 \tau _{12}^{\#} & 0 & 0 & 0 & 0 & 0 & 0 & 0 & 1-\frac{1}{2} \tau _{12} \tau _{12}^{\#} \\[5pt] 
\end{array}
\right]
\ee
Its  action on the separable purely ``bosonic'' state $\ket{00}$ gives
\be\label{eq:genstate1}
\begin{split}
\text{exp}(X_{\tau_{12}})\ket{00}=&\left(1+\frac{1}{2}\tau _{12} \tau _{12}^{\#}\right)\ket{00} -\tau _{12}^{\#}\ket{\bullet1},
\end{split}
\ee
where the sign on the final term follows from \eqref{eq:inoutsign}, which is a normalised entangled superposition of even and odd basis vectors.

The entangled state used to test Tsirelson's bound in \cite{Borsten:2012pp},
\be
\ket{\psi}=\left(1+\frac{1}{2}\tau^2+\frac{1}{2}\lambda^2+\frac{3}{4}\tau^2\lambda^2\right)\left[\alpha\ket{00}+\beta\ket{11}-\tau\ket{\bullet1}-\lambda\ket{1\bullet}\right]
\ee
where $\tau^2=\tau\tau^\#, \lambda^2=\lambda\lambda^\#$ and  $\alpha\alpha^\#+\beta\beta^\#=1$,  can be generated from $\ket{00}$ using a simple sequence of such transformations. This state  demonstrates that using $\UOSp(5|4)$ we can generate entangled states with non-vanishing superdeterminant \cite{Borsten:2009ae} starting from the separable $\ket{00}$ with vanishing superdeterminant. A simple set of elementary transformations of this type can also be used to generate the entangled state used in \cite{Bradler:2012ii}.

\section{Conclusions}

We have introduced the global super unitary group for $n$ superqubits. For $n=2$ we have seen that its bosonic subgroup is  transitive on the 2-qubit subspace, despite the fact it does not contain the usual 2-qubit unitary group $\SU(4)$. This argument used the transitive property of $\USp(4)$ on its 5-dimensional irreducible representation  spanned by $\{\ket{AB}, \ket{\bullet \bullet}\}$.   The appearance of the $\rep{5}$ of  $\USp(4)$, as opposed to the $\rep{4}$ encountered in the context of quantum control, is necessary since only the $\rep{5}$ correctly decomposes into the $(\rep{2,2})\oplus (\rep{1,1})$ under the 2-qubit local unitary group $\SU_A(2)\times\SU_B(2)$, allowing for the consistent  truncation from superqubits to  qubits.  Using this explicit example we have seen that, starting from a separable state, $\UOSp(5|4)$ can generate states with a non-vanishing entanglement measure, the superdeterminant \cite{Borsten:2009ae}. We will return to  entanglement measures and classification elsewhere.

Let us conclude, keeping  the 2-superqubit example in mind,  with some comments on three or more superqubits. For three qubits we have global super unitary group $\UOSp(14|13)$. The superqubits transform as a 
$
\rep{(14,1)\oplus (1,13)}
$
under the  bosonic subgroup $\USp(14)\times \SO(13)$. As for two superqubits, the 3-qubit global unitary group $\SU(8)$ is not contained in  $\USp(14)\times \SO(13)$, but the proper subgroup $\USp(8)\subset\SU(8)$ is. The even states branch according as  
\be
\begin{array}{cccccccc}
\USp(14)& \supset& \USp(8) &\times& \USp(6)\\
\rep{14}&\rightarrow & (\rep{8,1}) &\oplus & (\rep{1,6})
\end{array}
\ee
where the $\rep{8}$ is spanned by $\ket{ABC}$ and the $\rep{6}$ by  $\ket{A\bullet\bullet}, \ket{\bullet B\bullet}, \ket{\bullet\bullet C}$. Unlike two superqubits  the $\USp(8)$ here is equivalent to the $\USp(8)$ of quantum control, since in both cases we have an irreducible $\rep{8}$. This is consistent with the truncation to three qubits as
\be
\begin{array}{ccccccc}
\USp(8)& \supset& \SU(2)\times \SU(2)\times \SU(2)\\
\rep{8}&\rightarrow & (\rep{2,2,2})
\end{array}
\ee
It then follows immediately from \autoref{tab:spheres} that the bosonic subgroup acts transitively on the subspace of 3-qubit states using the same $\USp(8)\subset\SU(8)$ as discussed in \cite{Albertini:2003}.

Similarly,
\be
\begin{array}{ccccccc}
\USp(6)& \supset& \SU'(2)\times \SU'(2)\times \SU'(2)\\
\rep{6}&\rightarrow & (\rep{2,1,1})\oplus (\rep{1,2,1})\oplus (\rep{1,1, 2})
\end{array}
\ee 
so that there is a diagonal $\SU_A(2)\times \SU_B(2)\times \SU_C(2)\subset\USp(14)\times \SO(13)$ under which,
\be
\rep{14}\rightarrow \underbrace{(\rep{2,2,2})}_{\ket{ABC}} \oplus  \underbrace{(\rep{2,1,1})}_{\ket{A\bullet\bullet}}\oplus \underbrace{(\rep{1,2,1})}_{\ket{\bullet B\bullet}}\oplus \underbrace{(\rep{1,1, 2})}_{\ket{\bullet\bullet C}}
\ee
and, similarly for the odd $\rep{13}$ of $\SO(13)$, 
\be
\rep{13}\rightarrow  \underbrace{(\rep{2,2,1})}_{\ket{AB\bullet }}\oplus \underbrace{(\rep{2, 1, 2})}_{\ket{A\bullet C}}\oplus \underbrace{(\rep{1,2, 2})}_{\ket{\bullet B C}}\oplus \underbrace{(\rep{1,1,1})}_{\ket{\bullet\bullet \bullet}}
\ee 
making clear the structure of the three superqubit states with respect to the local unitaries inside $\USp(14|13)$.

Four superqubits is the first case for which the standard global unitary group $\SU(16)$ \emph{is} contained in the global super unitary group $\USp(41|40)$. The 4-qubit subspace spanned by $\ket{ABCD}$ transforms as the $\rep{16}$ of $\SU(16)$ and so question of transitivity does not appear. The standard $\SU(2^n)$ subgroup is present for all $n\geq 4$.

\section*{Acknowledgments}

This work was funded in part by a  Royal Society for an International Exchanges travel grant. We are grateful for the hospitality of  the Newton Institute, Cambridge and the Mathematical Institute, University of Oxford, where part of this work was done. The work of LB is supported by an Imperial College Junior Research Fellowship. The work of MJD is supported by the STFC under rolling grant ST/G000743/1.  

\appendix

\providecommand{\href}[2]{#2}\begingroup\raggedright\endgroup

\end{document}